\documentclass[10pt,twocolumn,letterpaper]{article}

\usepackage[pagenumbers]{wacv} 

\definecolor{wacvblue}{rgb}{0.21,0.49,0.74}
\usepackage[pagebackref,breaklinks,colorlinks,allcolors=wacvblue]{hyperref}
\usepackage{algorithm}
\usepackage{algpseudocode}
\usepackage{multirow}
\usepackage{arydshln}
\usepackage{pgfplots}
\pgfplotsset{compat=1.18}
\usepackage{xcolor}
\usepackage{tcolorbox}

\usepackage{amsmath, amssymb, amsthm}

\usepackage{amsmath,amssymb,amsthm}
\theoremstyle{plain}\newtheorem{theorem}{Theorem}
\newtheorem{lemma}{Lemma}
\newtheorem{proposition}{Proposition}
\theoremstyle{remark}
\theoremstyle{definition}

\def\wacvPaperID{457} 
\def\confName{WACV}
\def\confYear{2026}

\title{Distilling Diversity and Control in Diffusion Models}

\author{Rohit Gandikota$^{*}$ \hspace{5em} David Bau \hspace{.7em} \vspace{3pt} \\ 
Northeastern University}

\begin{document}
\twocolumn[{
\renewcommand\twocolumn[1][]{#1}
\maketitle
\includegraphics[width=\textwidth]{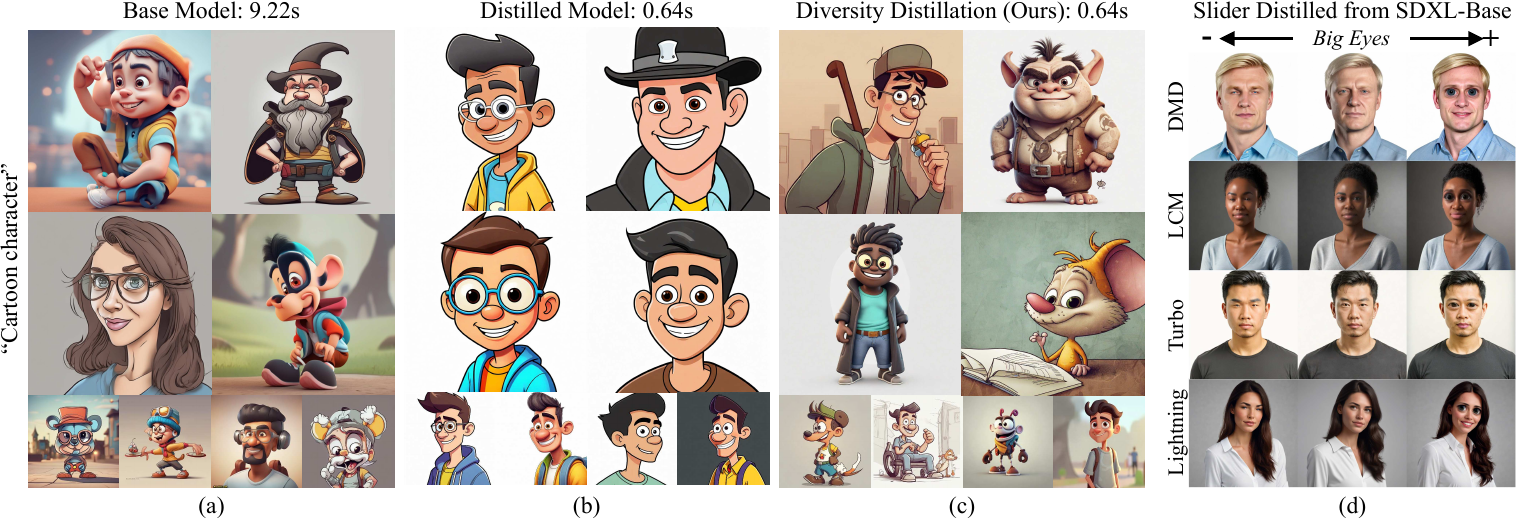}
\captionof{figure}{\textbf{Diversity Distillation:} (a) SDXL-Base is very slow (9.22s) but has good \emph{sample diversity} for the prompt "Cartoon character", sampling a wide range of styles, creatures, backgrounds, and poses. (b) SDXL-DMD2 is fast (0.64s) but sacrifices diversity. With the same prompt, samples all have the same style, pose, species, and context. (c) We show how the diversity of the base model can be distilled into the fast model by substituting the first timestep, achieving both speed and diversity (0.64s). \textbf{Control Distillation:} (d) Despite the lack of sample diversity in distilled models, control mechanisms like Concept Sliders trained on base models transfer perfectly to distilled variants, demonstrating that the representational structure for diversity exists but is not spontaneously activated during generation.}

\label{fig:intro}
\vspace{3.25mm}
}]
\begin{abstract}
Distilled diffusion models generate images in far fewer timesteps but suffer from reduced sample diversity when generating multiple outputs from the same prompt. To understand this phenomenon, we first investigate whether distillation damages concept representations by examining if the required diversity is properly learned. Surprisingly, distilled models retain the base model's representational structure: control mechanisms like Concept Sliders and LoRAs transfer seamlessly without retraining, and SliderSpace analysis reveals distilled models possess variational directions needed for diversity yet fail to activate them. This redirects our investigation to understanding how the generation dynamics differ between base and distilled models. Using $\hat{\mathbf{x}}_{0}$ trajectory visualization, we discover distilled models commit to their final image structure almost immediately at the first timestep, while base models distribute structural decisions across many steps. To test whether this first-step commitment causes the diversity loss, we introduce \textit{diversity distillation}, a hybrid approach using the base model for only the first critical timestep before switching to the distilled model. This single intervention restores sample diversity while maintaining computational efficiency. We provide both causal validation and theoretical support showing why the very first timestep concentrates the diversity bottleneck in distilled models. Our code and data are available at \href{https://distillation.baulab.info/}{\textcolor[rgb]{0.21,0.49,0.74}{distillation.baulab.info}}
\end{abstract}
\section{Introduction}
\label{sec:intro}

Distilled diffusion models generate images in far fewer timesteps but lack the sample diversity of their original base model counterparts. In this paper we ask: \emph{Why do distilled diffusion models collapse in sample diversity?}

\let\thefootnote\relax \footnote{$^{*}$Correspondence to \texttt{gandikota.ro@northeastern.edu}}

This limitation fundamentally constrains practical applications. When users generate multiple images from the same prompt, they expect varied outputs~\cite{artisticNeeds} that explore different creative interpretations. Figure~\ref{fig:intro} illustrates this problem: while base models produce diverse structural compositions across random seeds, distilled variants converge to visually similar results despite their computational advantages. This diversity collapse limits the utility of efficient models in creative workflows, design exploration, and applications requiring multiple candidate generations~\cite{artisticNeeds}.

The challenge is particularly puzzling given recent advances in distillation quality. Diffusion models demonstrate unprecedented generation quality~\cite{ho2020denoising,rombach2022high,podell2023sdxl,chen2023pixart,flux}, yet their computational demands create deployment barriers. Modern distillation techniques~\cite{dmd1,dmd,lcm,turbo,lightning,flux} have successfully maintained image quality while reducing inference steps from 50-100 to just 1-4. Some distilled models even achieve better distributional diversity than their base counterparts. However, distributional diversity is different from sample diversity. Distributional diversity is the ability to cover the full spectrum of training data across varied prompts, while sample diversity is the ability to generate diverse images for a single prompt under different seeds. SDXL-DMD2~\cite{dmd} shows superior distributional diversity with lower FID~\cite{heusel2017fid} scores on COCO~\cite{lin2014microsoft} datasets but exhibits poor sample diversity, making the collapse even more mysterious.

To solve this puzzle, we first investigate whether distillation damages the model's concept representations by analyzing the top variation directions~\cite{gandikota2025sliderspace} in base models and examining if these directions exist in distilled models. Surprisingly, we find that distilled models retain the variational directions needed for diversity—they simply fail to activate them during generation. This leads us to hypothesize that timestep dynamics could be the reason, supported by theoretical analysis showing how distillation collapses the output variance at early timesteps. We test this by visualizing what models predict at intermediate timesteps~\cite{ho2020denoising,wang2023diffusion}, revealing that distilled models commit to their final structure almost immediately at the first step while base models distribute decisions across many steps. To causally test our hypothesis, we introduce a hybrid inference technique and demonstrate that sample diversity can be drastically improved by simply modifying the first timestep of distilled models.

Our investigation reveals that the problem lies not in \emph{what} distilled models learn, but in \emph{how} they generate. Through careful analysis of model representations and generation dynamics, we identify the root cause and demonstrate a simple solution that restores sample diversity without sacrificing efficiency. Our findings challenge conventional assumptions about the diversity-efficiency tradeoff and provide actionable insights for both researchers developing distillation methods and practitioners deploying efficient diffusion models.

\section{Related Works}

\paragraph{Diffusion Distillation:} While diffusion models~\cite{sohl2015deep,ho2020denoising,song2020score} excel at high-quality image synthesis, their requirement for 20-100 sampling steps creates significant computational bottlenecks. Diffusion distillation techniques address this limitation by finetuning base models that maintain quality with fewer steps. Progressive distillation~\cite{salimans2022progressive} established the foundation by iteratively training student models to match teacher outputs with half the sampling steps. Recent approaches have further improved efficiency through distinct methodologies: Adversarial Diffusion Distillation~\cite{turbo}, implemented in SDXL-Turbo, integrates score distillation with adversarial training to enable high-fidelity generation in just 1-4 steps, effectively combining diffusion guidance with GAN-like discriminators. Distribution Matching Distillation ~\cite{dmd}, featured in SDXL-DMD2, takes a different approach by focusing on matching output distributions rather than specific trajectories, eliminating regression loss and implementing a two time-scale update rule that significantly improves training stability. For balancing quality and mode coverage, Progressive Adversarial Diffusion Distillation~\cite{lightning} in SDXL-Lightning employs staged training with specialized latent-space discriminators, offering flexibility through checkpoints optimized for 1-8 step inference. Latent Consistency Models~\cite{lcm}, applied in SDXL-LCM, ensure consistency in latent representations across noise levels for distillation, reducing steps to 4-8 while preserving generation quality. Despite these advances in efficiency, the relationship between model distillation and sample diversity has remained largely unexplored. 

\begin{figure*}
\centering
\includegraphics[width=\linewidth]{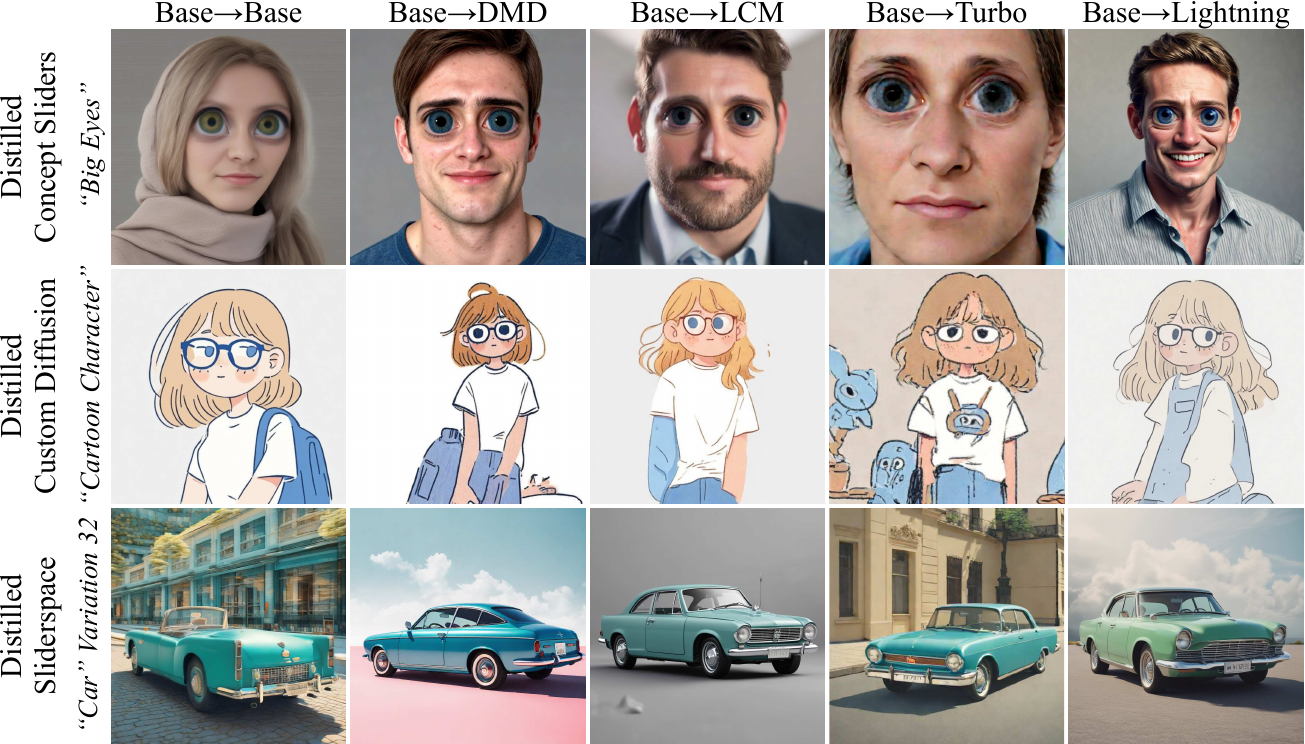}
\caption{Control directions (Sliders~\cite{kumari2023multi}), customization adapters (Custom Diffusion~\cite{gandikota2024concept}), and variational directions (Sliderspace~\cite{gandikota2025sliderspace}) trained on SDXL-Base transfer to all distilled models without additional finetuning. SliderSpace results suggest that top variation directions that capture sample diversity in the base model exist in distilled models but are not spontaneously activated during generation.}

\label{fig:transfer}
\end{figure*}


\textbf{Concept Representation:} Research in concept representation for diffusion models has evolved from basic personalization to sophisticated control mechanisms~\cite{zhang2023adding,ye2023ip,gandikota2024concept,mou2024t2i,cao2023masactrl,chefer2023attend}. Textual Inversion~\cite{gal2022image} captures the semantics of a concept with learnable embeddings in text space without modifying model weights, allowing personalization with just a few images. DreamBooth~\cite{ruiz2023dreambooth} advanced this approach by fine-tuning models with unique identifiers and a specialized prior preservation loss. Custom Diffusion~\cite{kumari2023multi} streamlined this process by optimizing only cross-attention layers, reducing storage requirements to just 3\% of model weights while enabling multi-concept customization simultaneously. For precise attribute manipulation, Concept Sliders~\cite{gandikota2024concept} introduced low-rank adaptors that create interpretable controls over specific visual attributes like age or weather conditions. This technique was expanded in SliderSpace~\cite{gandikota2025sliderspace}, which decomposes a model's visual variation capabilities, i.e. sample diversity, into multiple controls from a single prompt, enhancing creative exploration. Recent works have addressed the issue of suboptimal mode following in finetuned models by implementing an inference-time guidance annealing~\cite{jena2024elucidating}. Complementary to these control mechanisms, hierarchical concept trees~\cite{vinker2023concept,avrahami2023break} were developed to enable intuitive exploration of related visual concepts. Recent work has also addressed ethical concerns through targeted concept removal techniques by editing selective weights~\cite{gandikota2023erasing, gandikota2024unified, lu2024mace}, redirecting concept representations~\cite{kumari2023ablating, pham2024robust}. Since distillation modifies the UNet model of diffusion, in this work, we mainly focus on custom concept and control representations that are captured in UNet modules. Our work uniquely explores whether such control mechanisms can be distilled from base to efficient models without additional training.
\section{Control Distillation}
\label{sec:concept_distillation}

Having established that distilled models suffer from reduced sample diversity (Figure~\ref{fig:intro}), we investigate the underlying cause. Our first hypothesis is that distillation erodes the model's concept space: that the distilled model fails to learn or retain the representations necessary for diverse generation. If this were true, the model would naturally produce less varied outputs because the required conceptual building blocks would be missing or damaged.

To test this hypothesis, we examine whether the representational structure needed for diversity is preserved in distilled models. We pose this as \textit{representational compatibility}: do the variation directions and control mechanisms present in base models transfer to distilled models without retraining?

\begin{table*}[t]
\centering
\small
\resizebox{\linewidth}{!}{
\begin{tabular}{llc:cccc}
\textbf{Method} & \textbf{Concept} & \textbf{Base$\rightarrow$Base} & \textbf{Base$\rightarrow$DMD} & \textbf{Base$\rightarrow$LCM} & \textbf{Base$\rightarrow$Turbo} & \textbf{Base$\rightarrow$Lightning} \\
\hline
\multirow{3}{*}{Concept Sliders~\cite{gandikota2024concept}} & Age & 20.4 & 17.8 & 27.1 & 19.0 & 24.8 \\
& Smile  & 19.7 & 21.4 & 19.5 & 33.5 & 14.0 \\
& Muscular & 34.6 & 26.7 & 33.8 & 39.0 & 33.2 \\
\hdashline
\multirow{3}{*}{Customization~\cite{ruiz2023dreambooth, kumari2023multi}}  & Lego & 32.2 & 26.8 & 26.0 & 30.3 & 29.7 \\
& Watercolor style & 34.3 & 31.4 & 29.6 & 27.5 & 39.2 \\
 & Crayon style & 32.7 & 27.8 & 24.7 & 29.5 & 32.5 \\
\hdashline 

\multirow{3}{*}{\begin{tabular}[c]{@{}l@{}}Sliderspace~\cite{gandikota2025sliderspace}\\(Top n$^{\text{th}}$ Variation)\end{tabular}} & Direction 1 & 29.3 & 28.1 & 24.1 & 27.7 & 22.4 \\
                                                             & Direction 16 & 32.9 & 32.0 & 33.4 & 28.1 & 29.9 \\
                                                             & Direction 32 & 30.7 & 28.9 & 29.3 & 30.5 & 31.4 \\

\end{tabular}
}
\caption{We show the percentage change in CLIP score from the original image and the edited image. Higher values indicate stronger attribute change or style transfer. Control effectiveness is largely preserved when transferring from base to different distilled models, with only minor variations across distillation techniques. Importantly, SliderSpace directions for the concept ``car'' which capture the base's natural variational structure transfer. This demonstrates that the representational components needed for diversity exist in distilled models.
}
\label{tab:clip_similarity}
\end{table*}

\subsection{Experimental Setup}
\label{subsec:concept_setup}

We test three complementary families of controls that probe different aspects of model representations: 

\textbf{Concept Sliders}~\cite{gandikota2024concept,gandikota2025sliderspace} provide fine-grained control over visual attributes (e.g., age, weather, eye size) through low-rank adaptations. These test whether distilled models preserve the semantic understanding needed to manipulate specific visual properties.

\textbf{Customization mechanisms} including Custom Diffusion~\cite{kumari2023multi} and DreamBooth~\cite{ruiz2023dreambooth} capture nuanced, user-defined concepts through specialized training procedures. These mechanisms test whether distilled models retain the capacity to encode and recall complex, personalized concepts.

\textbf{SliderSpace}~\cite{gandikota2025sliderspace} decomposes the seed-induced variations of a base diffusion model into interpretable, continuous directions—each corresponding to a controllable visual factor (e.g., texture, lighting, layout). This directly analyzes whether the variational components responsible for sample diversity are present in distilled models.

For each mechanism, we perform bidirectional transfer experiments: training on base models and applying to distilled models, and vice versa. We study four SDXL distilled families—SDXL-Turbo~\cite{turbo}, SDXL-Lightning~\cite{lightning}, SDXL-LCM~\cite{lcm}, and SDXL-DMD2~\cite{dmd}—which span diverse distillation procedures. Additional training details and experiments on additional base models (SD 2.1~\cite{rombach2022high} and DiT-backboned PixArt~\cite{chen2023pixart}) are provided in the Appendix~\ref{sec:other_models}.

\subsection{Results}
\label{subsec:concept_results}

Our experiments reveal surprising evidence that contradicts our initial hypothesis. Figure~\ref{fig:transfer} demonstrates seamless transfer of control mechanisms across model variants. For example, a "comical big eyes" slider trained on SDXL effectively controls SDXL-Turbo's generations, despite Turbo requiring only 1-4 steps compared to SDXL's 20-100 steps.

Table~\ref{tab:clip_similarity} quantifies this compatibility through CLIP scores~\cite{hessel2021clipscore}. Transfer effectiveness remains consistently high across all tested combinations, with positive score changes indicating successful attribute manipulation. While absolute scores vary due to different guidance scales across models, the consistent positive changes confirm that concept representations are preserved during distillation. We show more experiments for the Distillation$\rightarrow$Base evidence in Appendix~\ref{sec:reverse}.

Most tellingly, SliderSpace~\cite{gandikota2025sliderspace} analysis reveals that variation directions learned on base models transfer perfectly to distilled variants without retraining (Tab~\ref{tab:clip_similarity} and Fig.~\ref{fig:transfer}). This indicates that the factors needed for diverse generation are present and accessible in distilled models. However, during free generation with different random seeds, distilled models rarely activate these factors.

These results directly contradict our initial hypothesis that distillation destroys concept representations. The concepts remain encoded and controllable in distilled models; what fails is their spontaneous activation during few-step sampling. This finding redirects our investigation from \emph{what is missing} to \emph{why existing diversity mechanisms fail to activate}. If the required variational directions exist but are not used, the problem must lie in the generation dynamics rather than the representational capacity.
\section{\texorpdfstring{$\hat{\mathbf{x}}_{0}$}{x̂₀}: Visualizing Intermediate Latents}
\label{sec:dtviz}

\begin{figure}[t]
    \centering
    \includegraphics[width=\linewidth]{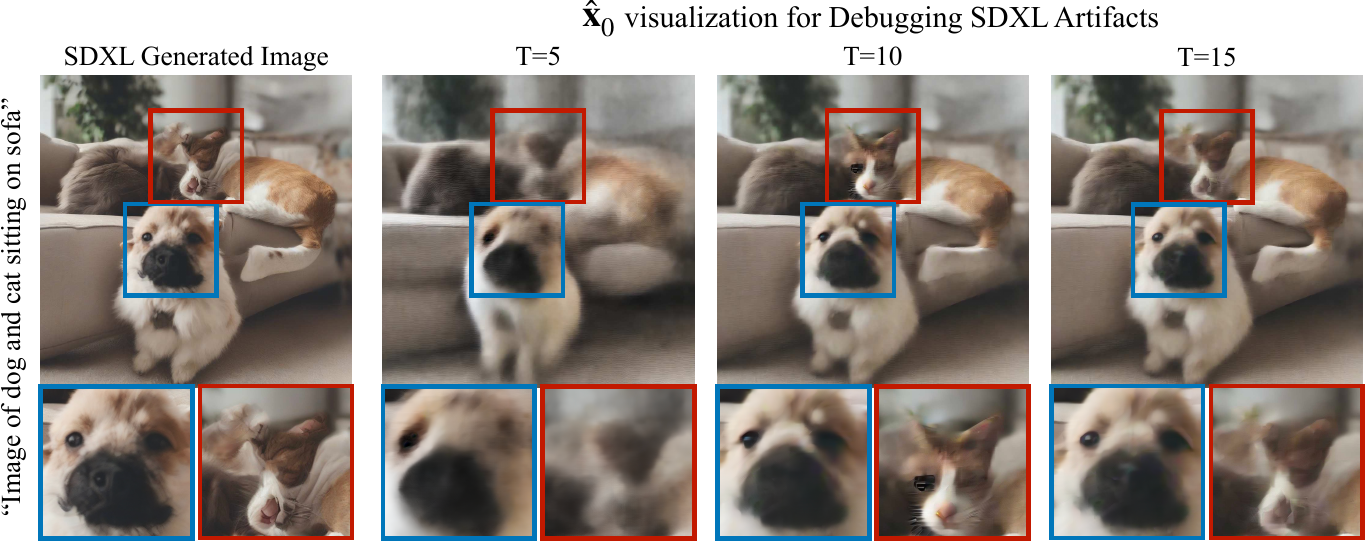}
    \caption{$\hat{\mathbf{x}}_{0}$ visualization reveals generation inconsistencies. When prompted with ``Image of dog and cat sitting on sofa,'' the SDXL model produces an image with only a dog. However, $\hat{\mathbf{x}}_{0}$ visualization at $T=10$ shows the model initially conceptualizing a cat face (red box) before abandoning this element in the final generation. This demonstrates how diffusion models can discard semantic elements during the denoising process.}
    \label{fig:dtdebug}
\end{figure}

Having established that distilled models possess the representational components needed for diversity, we must understand why these components fail to activate during generation. Our investigation shifts from \emph{what} distilled models learn to \emph{how} they generate. To analyze the generation dynamics, we employ $\hat{\mathbf{x}}_{0}$ trajectory visualization as a diagnostic tool.

In the standard diffusion formulation~\cite{ho2020denoising,sohl2015deep}, $\mathbf{x}_0$ represents the clean image that is progressively corrupted with noise over $T$ timesteps. During reverse generation, at each timestep $t$, the model predicts noise $\epsilon_\theta(\mathbf{x}_t, t)$ to compute the next denoising step:

Let $\mathbf{x}_0$ be an initial image and $\mathbf{x}_T$ be pure Gaussian noise. The forward diffusion process gradually adds noise according to a variance schedule $\beta_t$, with corresponding noise level parameters $\alpha_t = 1 - \beta_t$ and cumulative parameters $\bar{\alpha}_t = \prod_{s=1}^{t} \alpha_s$. The generative process aims to reverse this diffusion, starting from $\mathbf{x}_T$ and progressively denoising to reconstruct $\mathbf{x}_0$. At timestep $t$, the model predicts noise $\epsilon_\theta(\mathbf{x}_t, t)$ to compute the next step:

\begin{figure*}[t]
\centering
\includegraphics[width=\linewidth]{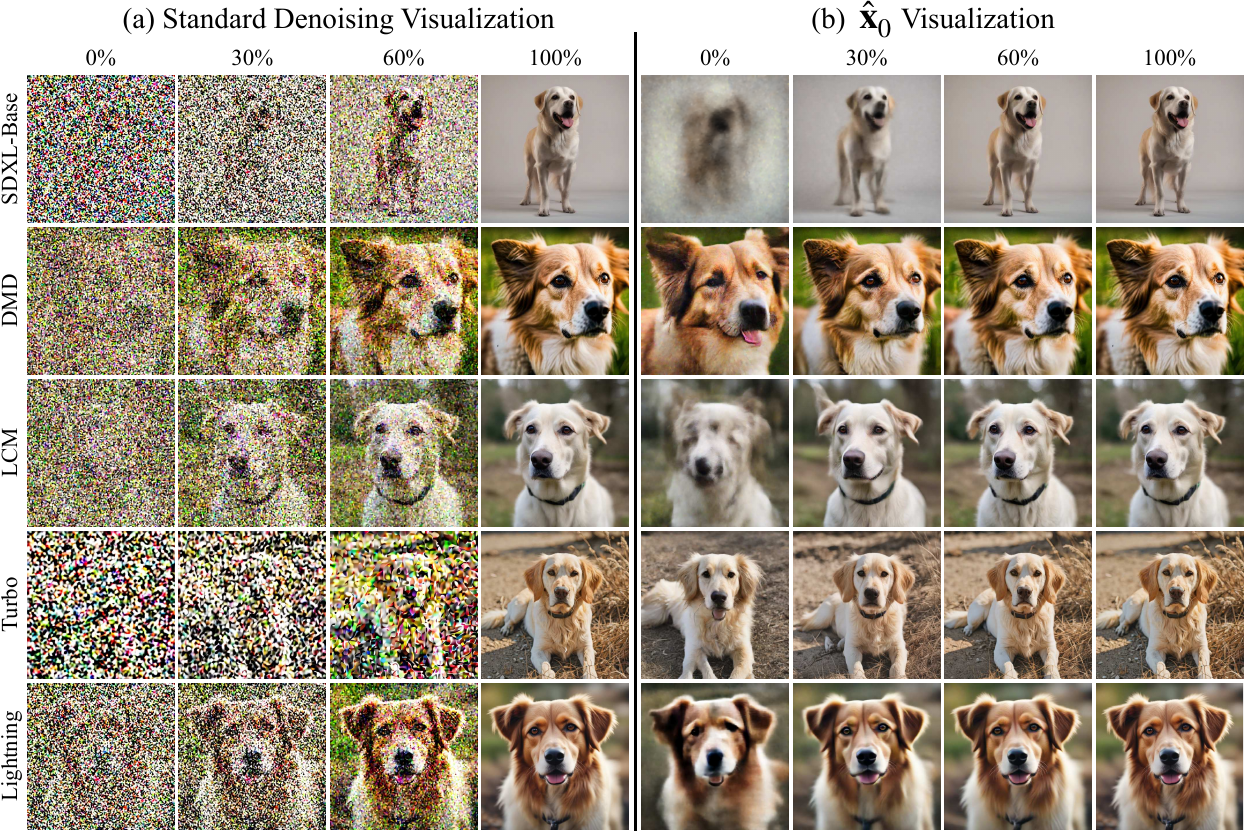}
\caption{Comparison of standard diffusion visualization vs. $\hat{\mathbf{x}}_{0}$ visualization. (a) Standard visualization of intermediate latents shows subtle differences between base and distilled models. (b) $\hat{\mathbf{x}}_{0}$ visualization reveals dramatic differences in how models predict the final output. Distilled models commit to final image structure in the first timestep, while base models gradually refine structure across multiple steps, explaining the observed mode collapse in distilled models.}
\label{fig:extrapolation}
\end{figure*}

\begin{equation}
\label{eq:xt}
\mathbf{x}_{t-1} = \frac{\mathbf{x}_t - \sqrt{1-\alpha_t}\epsilon_\theta(\mathbf{x}_t, t)}{\sqrt{\alpha_t}}
\end{equation}

Using this predicted noise $\epsilon_\theta$, we can estimate what the model believes the final clean image will be at any intermediate timestep $t$, by extrapolating the same noise prediction all the way to $\mathbf{x}_0$:

\begin{equation}
\label{eq:x0t}
\hat{\mathbf{x}}_{0|t} = \frac{\mathbf{x}_t - \sqrt{1-\bar{\alpha}_t}\epsilon_\theta(\mathbf{x}_t, t)}{\sqrt{\bar{\alpha}_t}}
\end{equation}
Visualizing these $\hat{\mathbf{x}}_{0|t}$ trajectories across timesteps reveals when structural decisions are made and how they evolve during generation~\cite{wang2023diffusion}. This technique allows us to peer into the model's "thought process" and understand the temporal dynamics of image formation.

\subsection{\texorpdfstring{$\hat{\mathbf{x}}_{0}$}{x̂₀} for Investigating Generation Artifacts}

Before applying this technique to our diversity investigation, we demonstrate its utility for understanding generation artifacts. In Figure~\ref{fig:dtdebug}, when prompted with "Image of dog and cat sitting on sofa," the SDXL model produces a final image containing only a dog's face. However, $\hat{\mathbf{x}}_{0|t}$ visualization at early timesteps ($T=10$) reveals that the model initially conceptualized a cat face (highlighted in red box) before abandoning this element in later steps. This insight exposes how diffusion models can "change their mind" during generation, sometimes discarding semantic elements present in the prompt.

\subsection{\texorpdfstring{$\hat{\mathbf{x}}_{0}$}{x̂₀} for Investigating Sample Diversity}

We now apply $\hat{\mathbf{x}}_{0}$ visualization to compare base and distilled models under identical conditions: same seed and prompt ("image of a dog"). Standard visualizations of intermediate latents $\mathbf{x}_t$ (Fig~\ref{fig:extrapolation}.a) show only subtle differences across base and distilled models. In stark contrast, the $\hat{\mathbf{x}}_{0}$ trajectories (Fig~\ref{fig:extrapolation}.b) expose a dramatic pattern that could potentially explain the diversity collapse.

Distilled models commit to their final image structure almost immediately after the first timestep. Their $\hat{\mathbf{x}}_{0|t}$ predictions quickly converge to the final output. Base models, conversely, distribute structural decision-making across many timesteps, with their $\hat{\mathbf{x}}_{0|t}$ predictions gradually refining and evolving toward the final image. 

\begin{figure}
\centering
\includegraphics[width=\linewidth]{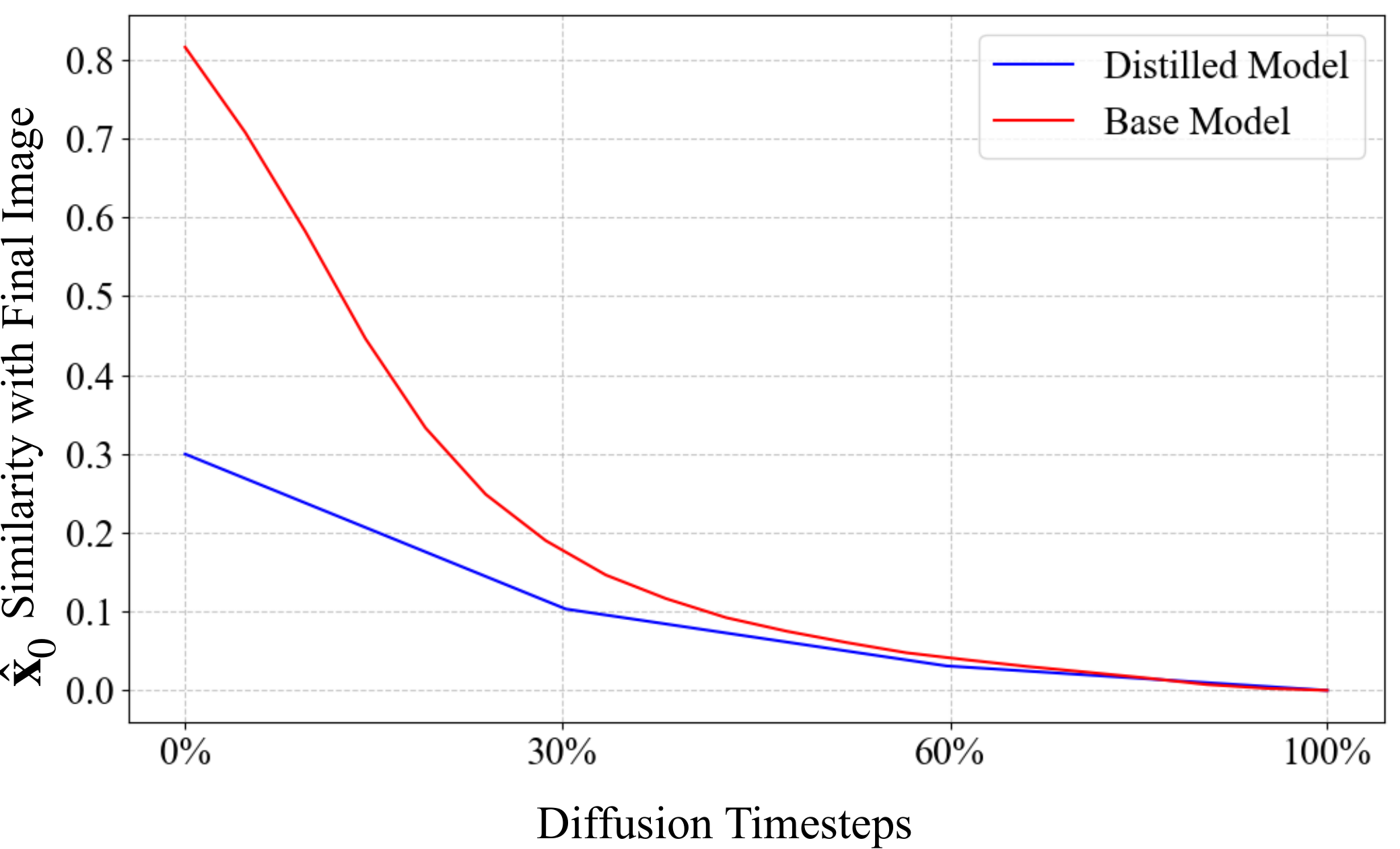}
\caption{Measuring the dreamsim distance between intermediate $\hat{\mathbf{x}}_{0}$ visualization and final generated image reveals that distilled models establish structural image composition within the initial diffusion step, whereas base models require approximately 30\% of steps to achieve comparable structural definition.}
\label{fig:dtdistance}
\end{figure}

Figure~\ref{fig:dtdistance} quantifies this phenomenon by measuring DreamSim distances~\cite{fu2023dreamsim} between intermediate $\hat{\mathbf{x}}_{0|t}$ predictions and final outputs across COCO-10k~\cite{lin2014microsoft} prompts. The data reveals that distilled models achieve a large fraction of their final structure after a single timestep, while base models require approximately 30\% of their total inference steps to reach comparable structural definition. We provide more qualitative evidence in Appendix~\ref{sec:more_x0}.

Our discovery suggests a testable hypothesis: the first timestep is the primary culprit behind diversity loss in distilled models. We provide theoretical analysis (detailed in Appendix~\ref{sec:theory}) showing why the first timestep concentrates the diversity bottleneck. The analysis demonstrates how timestep compression in distillation amplifies decision-making pressure early in the process, with the amplification factor being largest at initial timesteps.

\section{Diversity Distillation}
\label{sec:diversity_distillation}
\begin{figure*}
\centering
\includegraphics[width=\linewidth]{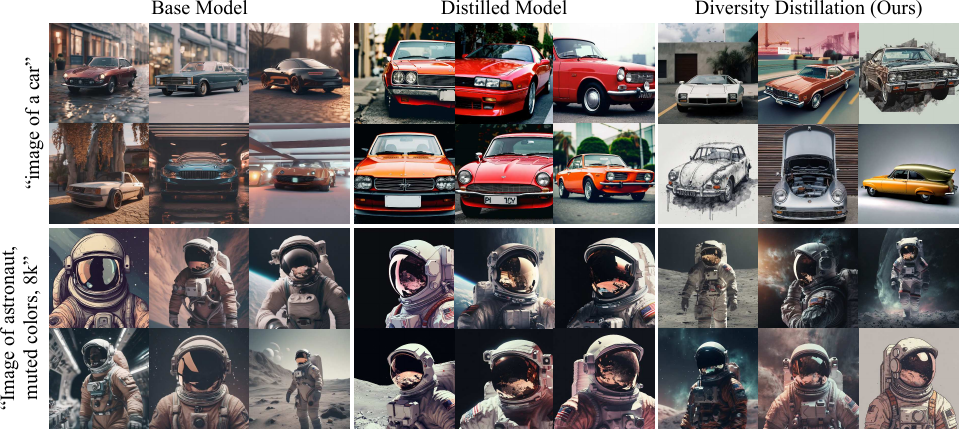}
\caption{\textbf{Visual comparison of generation diversity.} Each row shows three different generations (different random seeds) for the same prompt using: (left) base model, (middle) distilled model, and (right) our diversity distillation approach. Note how the distilled model produces similar car types, poses, and astronaut orientations across seeds, while our approach restores diversity in colors, contexts, and compositions comparable to the base model while maintaining similar inference speed.}
\label{fig:modecollapse}
\end{figure*}
Our $\hat{\mathbf{x}}_{0}$ analysis identified a clear hypothesis: the first timestep is the primary bottleneck causing diversity collapse in distilled models. To test this causally, we design experiments that directly manipulate the first timestep and measure the impact on sample diversity. If our hypothesis is valid, targeted interventions at this critical step should restore diversity without affecting efficiency.

We develop two approaches to test this hypothesis. First, we replace the first timestep of distilled model inference with the corresponding step from the base model. Second, we examine whether simply skipping the problematic first timestep can improve diversity. Both approaches target the identified bottleneck while preserving the computational advantages of distilled models.

Algorithm~\ref{alg:diverse} implements our hybrid inference strategy. The approach uses the base model for the critical first timestep(s) to establish diverse structural foundations, then transitions to the distilled model for efficient completion. This design directly tests whether the first timestep controls diversity while maintaining computational efficiency.

\begin{algorithm}
\caption{Hybrid Inference for Diversity Distillation}
\label{alg:diverse}
\begin{algorithmic}[1]
\Require Base model $f_{\text{base}}$, distilled model $f_{\text{distil}}$, total timesteps $T$, transition point $k$
\Ensure Generated image $\mathbf{x}_0$
\State Initialize $\mathbf{x}_T \sim \mathcal{N}(0, \mathbf{I})$
\For{$t = T, T-1, \ldots, 1$}
    \If{$t > T-k$} \Comment{Critical timesteps for diversity}
        \State $\mathbf{x}_{t-1} \gets f_{\text{base}}(\mathbf{x}_t, t, \text{prompt})$ 
    \Else \Comment{Efficient refinement timesteps}
        \State $\mathbf{x}_{t-1} \gets f_{\text{distil}}(\mathbf{x}_t, t, \text{prompt})$ 
    \EndIf
\EndFor
\State \Return $\mathbf{x}_0$
\end{algorithmic}
\end{algorithm}

\subsection{Experimental Results}
We evaluate our approach across two complementary diversity metrics that capture different aspects of model performance. \textbf{Distributional diversity} measures how well generated samples match the real training data distribution across varied prompts, assessed through FID~\cite{heusel2017fid} (lower is better). \textbf{Sample diversity} measures variation among outputs generated from the same prompt with different random seeds, quantified by average pairwise DreamSim distance~\cite{fu2023dreamsim} (higher is better).

\paragraph{Distributional Diversity.} Table~\ref{tab:comparison} demonstrates that our hybrid approach achieves superior distributional diversity compared to both base and distilled models. Measuring FID against the COCO-30k dataset, our method achieves better scores than the base model while maintaining the computational efficiency of distilled inference. This confirms that our intervention does not compromise the model's ability to handle diverse prompts. We hypothesize the improvement in FID compared to the base model is due to the usage of the distilled model in final timesteps. Prior work suggests that varied guidance levels in final timesteps can affect FID metrics~\cite{kynkaanniemi2024applying}.

\paragraph{Sample Diversity.} Table~\ref{tab:sample_diversity} reveals that our approach dramatically restores sample-level diversity. For each prompt, we generate 100 images with different random seeds and calculate average pairwise DreamSim distances. The hybrid approach restores the diversity lost during distillation.

\begin{table}
\centering
\small
\resizebox{\linewidth}{!}{
\begin{tabular}{lccc}

Prompt & Base & Distilled & Hybrid (Ours) \\
\hline
Sunset beach & \textbf{0.396} & 0.271 & 0.373 \\
Cute puppy & 0.233 & 0.199 & \textbf{0.265} \\
Futuristic city & 0.237 & 0.198 & \textbf{0.283} \\
Person & \textbf{0.484} & 0.347 & 0.461 \\
Van Gogh art & 0.337 & 0.305  & \textbf{0.366} \\
\hdashline
Average & 0.337 & 0.264 & \textbf{0.350} \\

\end{tabular}
}
\caption{Sample diversity measured by average pairwise DreamSim distance (higher is more diverse). Our hybrid approach restores the lost diversity in distillation with a simple intervention.}
\label{tab:sample_diversity}
\end{table}

Figure~\ref{fig:modecollapse} provides visual confirmation of these quantitative results. The distilled model clearly exhibits reduced structural variety across random seeds, producing similar compositions and layouts. Our hybrid approach successfully restores this diversity, generating varied structural arrangements comparable to the base model while maintaining fast inference speeds.

\paragraph{Causal Evidence.} To causally close the loop, we also provide an experiment in the Appendix~\ref{sec:causal} where keeping the first timestep with the distilled model and replacing the later timesteps with the base model does not help with improvement in sample diversity. This demonstrates that the first timestep is indeed the critical bottleneck, validating our mechanistic understanding derived from $\hat{\mathbf{x}}_{0}$ visualization.

\begin{figure*}[h]
\centering
\includegraphics[width=\linewidth]{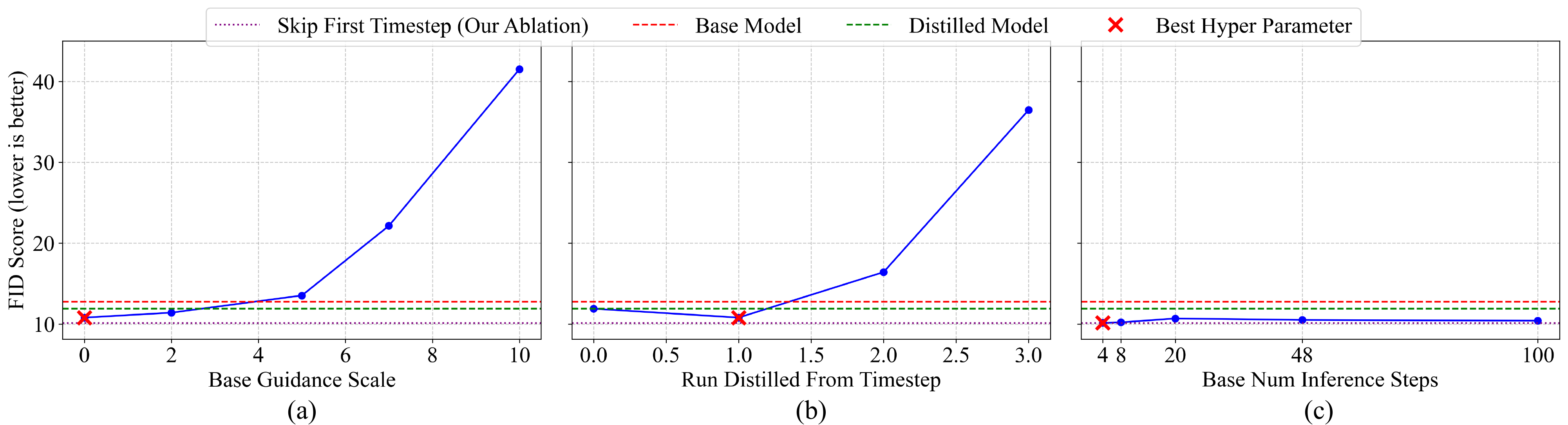}
\caption{  (a) Impact of guidance scale from the base model on diversity shows optimal performance around 0 guidance. (b) Effect of the number of distilled model steps ($k$) being replaced by base model inference. Running distilled model from first timestep ($k = 1$) provides diversity gains with minimal computational overhead. (c) Comparing the total timesteps of base model when replacing the first timestep of distilled model shows that replacing 1-1 timesteps of distilled with base is most ideal.}
\label{fig:hyperparam}
\end{figure*}
\begin{table}[t]
\centering
\small
\resizebox{\linewidth}{!}{
\begin{tabular}{rccccc}
\textbf{Method} & \textbf{Steps} & \textbf{FID($\downarrow$)} & \textbf{IS($\uparrow$)} & \textbf{CLIP($\uparrow$)} & \textbf{Time (s)($\downarrow$)} \\
\midrule
Base & 50 & 12.74 & 24.74 & 31.83 & 9.22 \\
Distilled & 4 & 15.52 & \textbf{27.20} & 31.69 & \textbf{0.64} \\
Hybrid (Ours) & 4 & \textbf{10.79} & 26.13 & \textbf{32.12} & \textbf{0.64} \\
Skip First Timestep & 3 & \textbf{10.12} & 24.69 & 31.71 & \textbf{0.53} \\
\end{tabular}
}
\caption{Measuring distributional diversity using FID shows our approach achieves comparable or better diversity than SDXL-Base while maintaining SDXL-DMD~\cite{dmd} efficiency. While our focus is sample diversity (Table~\ref{tab:sample_diversity}), this confirms our intervention preserves dataset coverage. Skipping first timestep achieves superior FID but lower generative quality (CLIP~\cite{hessel2021clipscore}, IS~\cite{salimans2016IS}).}
\label{tab:comparison}
\end{table}


\subsection{Hyperparameter Analysis and Efficiency}
Figure~\ref{fig:hyperparam} analyzes the key parameters of our approach. Most importantly, using the base model for just the first timestep ($k$=1) provides substantial diversity gains with minimal computational overhead (Fig~\ref{fig:hyperparam}.b), confirming our hypothesis about first-step criticality. The guidance scale analysis shows optimal performance around zero guidance from the base model, suggesting that natural diversity is preserved without additional steering(Fig~\ref{fig:hyperparam}.a). Finally, setting the base model's noise schedule similar to distilled (n=4) has optimal results (Fig~\ref{fig:hyperparam}.c). We find that using exact inference conditions for distilled models where the model weights are swapped with base for the first timestep is ideal.

For scenarios where loading both models simultaneously is not feasible, we explore skipping the first timestep entirely. Table~\ref{tab:comparison} shows this approach provides significant diversity improvements, though our hybrid method achieves superior quality as measured by CLIP and Inception scores. We provide qualitative examples in the Appendix~\ref{sec:more_examples}.

Our hybrid method demonstrates that the conventional efficiency-diversity tradeoff in distilled models can be resolved through targeted intervention. By identifying and addressing the specific timestep responsible for diversity collapse, we restore sample variety without sacrificing computational advantages. Our theoretical support in Appendix~\ref{sec:theory} provides a principled explanation for our empirical observations and suggests that distributing first-timestep decisions across multiple steps during training could be a promising direction for future distillation methods.
\section{Limitations}
While our approach significantly improves diversity without substantial computational overhead, some limitations remain. First, our method requires loading both base and distilled models in memory, increasing resource requirements compared to traditional inference. Future distillation work could explore our insights to design diversity-preserving mechanisms directly into a distilled model.

Second, our analysis focuses primarily on visual diversity metrics. Further investigation is needed to understand the impact on semantic diversity: the range of concepts and compositions a model can generate. Developing more diversity metrics that capture both visual and semantic variations can provide deeper insights into distillation.

Finally, our approach treats all prompts uniformly, but different concepts may benefit from different base/distilled step allocations. Adaptive inference strategies that dynamically adjust the transition point based on prompts could further optimize the quality-efficiency trade-off.

\section{Conclusion}
This work addresses a fundamental limitation of distilled diffusion models: the trade-off between computational efficiency and sample diversity. Our contributions are threefold: (1) We demonstrate that distilled models retain all the variational directions needed for diversity, contradicting the hypothesis that distillation damages concept representations—the required diversity mechanisms exist but fail to activate during generation; (2) We identify the root cause using $\hat{\mathbf{x}}_{0}$ trajectory visualization, revealing that distilled models concentrate structural decision-making in the first timestep while base models distribute decisions across many steps, and provide theoretical support showing why the first timestep creates a diversity bottleneck; and (3) We validate this theory by developing diversity distillation, demonstrating that targeted intervention at a single timestep restores full base model sample diversity as measured by DreamSim scores with strong empirical validation.

Our experimental results challenge the conventional diversity-efficiency trade-off. Diversity distillation restores the diversity of the original base model while maintaining the computational efficiency of distilled inference (\texttt{0.64s} vs. \texttt{9.22s} per image). By providing both mechanistic understanding and theoretical grounding for why diversity collapse occurs, our approach eliminates this traditional trade-off without additional training or model modifications, opening new possibilities for deploying efficient yet diverse generative models in creative applications.
\section*{Acknowledgment}
RG and DB are supported by Open Philanthropy and NSF grant \#2403304.

\section*{Code}
Our methods are available as open-source code. Source code, and data sets for reproducing our results can be found at \href{https://distillation.baulab.info/}{\textcolor[rgb]{0.21,0.49,0.74}{distillation.baulab.info}} and at our GitHub repo \href{https://github.com/rohitgandikota/distillation/}{\textcolor[rgb]{0.21,0.49,0.74}{github.com/rohitgandikota/distillation}}

{
    \small
    \bibliographystyle{ieeenat_fullname}
    \bibliography{main}
}
\clearpage
\appendix
\setcounter{page}{1}
\counterwithin{figure}{section}
\counterwithin{table}{section}
\counterwithin{equation}{section}

\maketitlesupplementary

\section{Theoretical Support}
\label{sec:theory}
We provide a theoretical justification for the empirical observation that distilled diffusion models lose most of their sample diversity at the \emph{first denoising timestep}. Our argument relies on the standard DDPM forward process, the amplification structure of $\hat{x}_{0|t}$, and the statistical effect of mean-squared-error (MSE) based distillation.

\subsection{Preliminaries}

Let $x_0 \in \mathbb{R}^d$ be a clean data sample, and consider the forward diffusion process
\begin{equation}
    q(x_t \mid x_0) = \mathcal{N}\!\left(x_t; \sqrt{\bar\alpha_t}\,x_0,\, (1-\bar\alpha_t)\mathbf{I}\right),
\end{equation}
where $\bar\alpha_t = \prod_{s=1}^t \alpha_s$ and $\alpha_s \in (0,1)$. 

The reverse model is often parameterized by predicting the noise $\varepsilon_\theta(x_t,t)$. From the forward relation, one can form an estimator of $x_0$:
\begin{equation}
    \hat{x}_{0|t} \;=\; \frac{x_t - \sqrt{1-\bar\alpha_t}\,\varepsilon_\theta(x_t,t)}{\sqrt{\bar\alpha_t}}.
    \label{eq:x0-hat}
\end{equation}

Equation~\eqref{eq:x0-hat} will be central to our analysis, as it determines how prediction errors or randomness in $\varepsilon_\theta$ propagate into variability in $\hat{x}_{0|t}$.

\subsection{Sensitivity}

\begin{lemma}[Sensitivity]
Let $\Delta \varepsilon$ denote a perturbation in the noise prediction at timestep $t$. Then the induced change in $\hat{x}_{0|t}$ is
\begin{equation}
    \Delta \hat{x}_{0|t} \;=\; -\sqrt{\tfrac{1-\bar\alpha_t}{\bar\alpha_t}} \,\Delta \varepsilon.
\end{equation}
\end{lemma}

\begin{proof}
Differentiate \eqref{eq:x0-hat} with respect to $\varepsilon$:
\[
\frac{\partial \hat{x}_{0|t}}{\partial \varepsilon} \;=\; -\frac{\sqrt{1-\bar\alpha_t}}{\sqrt{\bar\alpha_t}} I.
\]
Multiplying by $\Delta \varepsilon$ yields the result.
\end{proof}

\noindent The amplification factor
\begin{equation}
    \label{eq:amp}
    A_t \;=\; \sqrt{\frac{1-\bar\alpha_t}{\bar\alpha_t}}
\end{equation}
quantifies how strongly prediction variability at timestep $t$ is magnified in the clean-sample estimate. Since $\bar\alpha_t \ll 1$ at early timesteps, $A_t$ is very large.

\subsection{Distillation and Conditional Variance}

Distilled diffusion models are typically trained to minimize an MSE-style loss between student and teacher outputs. A standard fact from estimation theory is that the MSE minimizer of a random target is the conditional mean:
\begin{equation}
    s^*(x) = \mathbb{E}[Y \mid X=x].
\end{equation}
Thus, when the teacher output $Y$ has conditional variance $\operatorname{Var}(Y \mid X)$, the student collapses this variance and learns to reconstruct the mean.

By the law of total variance,
\begin{equation}
    \operatorname{Var}(Y) = \operatorname{Var}(\mathbb{E}[Y\mid X]) + \mathbb{E}\!\left[ \operatorname{Var}(Y\mid X)\right].
\end{equation}

MSE distillation removes the second term, reducing sample diversity by precisely $\mathbb{E}[ \operatorname{Var}(Y\mid X)]$.
\begin{equation}
   \Delta \operatorname{Var} = \operatorname{Var}(Y) - \operatorname{Var}(s^*(x)) = \mathbb{E}\!\left[ \operatorname{Var}(Y\mid X)\right].
\end{equation}

\subsection{Amplification of Diversity Loss at Early Timesteps}

Let the teacher produce a stochastic noise prediction $\varepsilon_{\mathrm{T}}(x_t,t;\xi)$, where $\xi$ captures randomness in sampling. Then, from Lemma 1,
\begin{equation}
    \operatorname{Var}(\hat{x}_{0|t} \mid x_t) 
    \;=\; \frac{1-\bar\alpha_t}{\bar\alpha_t} \; \operatorname{Var}(\varepsilon_{\mathrm{T}}(x_t,t)\mid x_t).
\end{equation}

Taking expectation over $x_t$,
\begin{equation}
    \mathbb{E}_{x_t}\big[\operatorname{Var}(\hat{x}_{0|t} \mid x_t)\big]
    \;=\; \frac{1-\bar\alpha_t}{\bar\alpha_t}\;\mathbb{E}_{x_t}\big[\operatorname{Var}(\varepsilon_{\mathrm{T}}(x_t,t)\mid x_t)\big].
    \label{eq:variance-amplification}
\end{equation}

\begin{proposition}[Amplified Diversity Loss]
For an MSE-trained student, the reduction in total variance of $\hat{x}_{0|t}$ due to distillation satisfies
\begin{equation}
    \Delta \operatorname{Var} \;\geq\; \frac{1-\bar\alpha_t}{\bar\alpha_t}\;\mathbb{E}_{x_t}\!\big[\operatorname{Var}(\varepsilon_{\mathrm{T}}(x_t,t)\mid x_t)\big].
\end{equation}
\end{proposition}

\begin{proof}
Direct application of the law of total variance to the random variable $\hat{x}_{0|t}$, substituting from \eqref{eq:variance-amplification}.
\end{proof}

\subsection{Main Result}

\begin{theorem}[First-Timestep Dominance]
Let $\bar\alpha_t$ denote the cumulative product of noise schedule parameters. Then for small $\bar\alpha_t$ (early timesteps), the amplification factor $(1-\bar\alpha_t)/\bar\alpha_t$ is maximal. Consequently, the diversity loss $\Delta \operatorname{Var}$ induced by MSE-based distillation is largest at the earliest timesteps. In particular, the first denoising step dominates the reduction of sample diversity in distilled diffusion models.
\end{theorem}

\begin{proof}
Since $\bar\alpha_t$ is monotonically increasing in $t$ with $\bar\alpha_0 \approx 0$, the fraction $(1-\bar\alpha_t)/\bar\alpha_t$ is strictly decreasing in $t$. Hence the bound on $\Delta\operatorname{Var}$ is largest at the earliest timestep, completing the proof.
\end{proof}

This proof applies in cases in which the teacher uses a stochastic $\epsilon_T(x_t, t; \xi)$  to create variance in its output.  This is the case for the ADD distillation method~\cite{turbo} that is used to train SDXL-Turbo, which uses stochastically sampled teacher for reconstruction loss. The variance trade-off in such cases is also analyzed in previous works~\cite{guo2024smooth}.

We have also tested our methods on models that do not use a stochastic teacher, such as DMD~\cite{dmd1}, which uses the Distribution Matching distillation method which uses a deterministic teacher, in which the teacher's $\operatorname{Var}(Y_t\mid x_t,t)=0$.  Despite this difference, we still measure a large drop in sample diversity concentrated at the first timestep.  We hypothesize that in these cases, the loss in diversity is due to sparse sampling of the teacher relative to the large diversity of text prompts, which allows the student to collapse to sparsely-sampled modes in a similar way as seen in the stochastic teachers, despite the deterministic teachers' theoretical diversity: for example in DMD only 100k text-conditioned teacher samples are used.  The loss in diversity due to sparse sampling of a teacher would be amplified in the early timesteps due to Eq~\eqref{eq:amp}.



\section{Control Distillation: Reverse Transfer}
\label{sec:reverse}
In the main paper, we demonstrated that control mechanisms trained on base models can be seamlessly transferred to distilled models. Here, we present additional results for the reverse direction: transferring control mechanisms trained on distilled models to base models. This bidirectional transfer capability further validates our hypothesis that concept representations are preserved during the distillation process.

\begin{figure*}
\centering
\includegraphics[width=\textwidth]{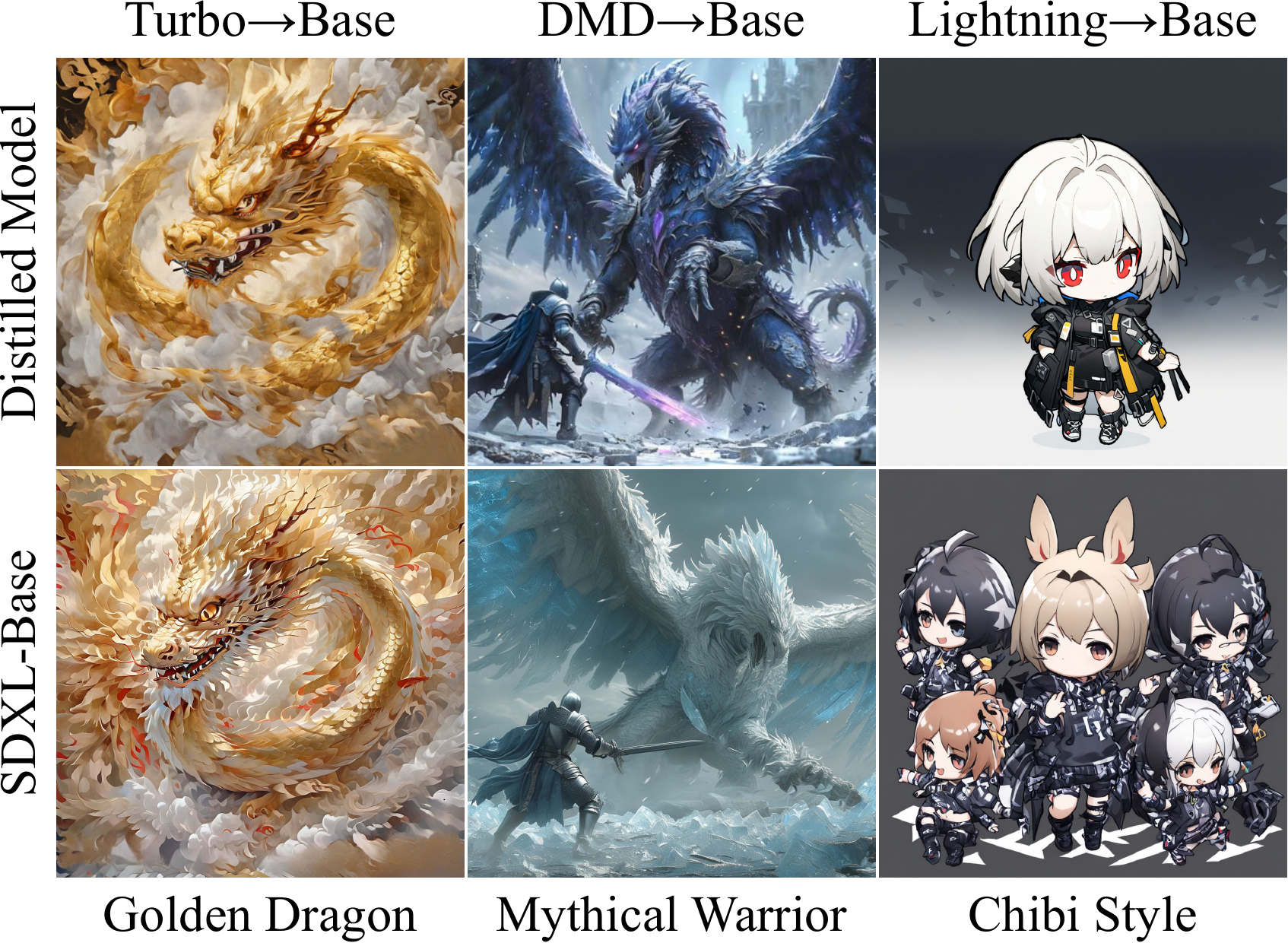}
\caption{Reverse Control Transfer: Control mechanisms (Custom Diffusion \cite{kumari2023multi} and Concept Sliders \cite{gandikota2024concept}) trained on distilled models can be effectively transferred to base models without retraining. This bidirectional transferability confirms that concept representations are preserved during diffusion distillation. Note: LCM LoRA transfers were excluded due to training difficulties with the LCM architecture.}
\label{fig:reverse_transfer}
\end{figure*}

We note that while most control mechanisms transferred effectively, we encountered difficulties training LoRA adaptations on LCM due to its specialized architecture and training procedure. These challenges highlight potential avenues for future research in developing more universally transferable control mechanisms.

\section{Skip Step Approach}
\label{sec:skip_first_step}
\begin{figure*}
\centering
\includegraphics[width=\textwidth]{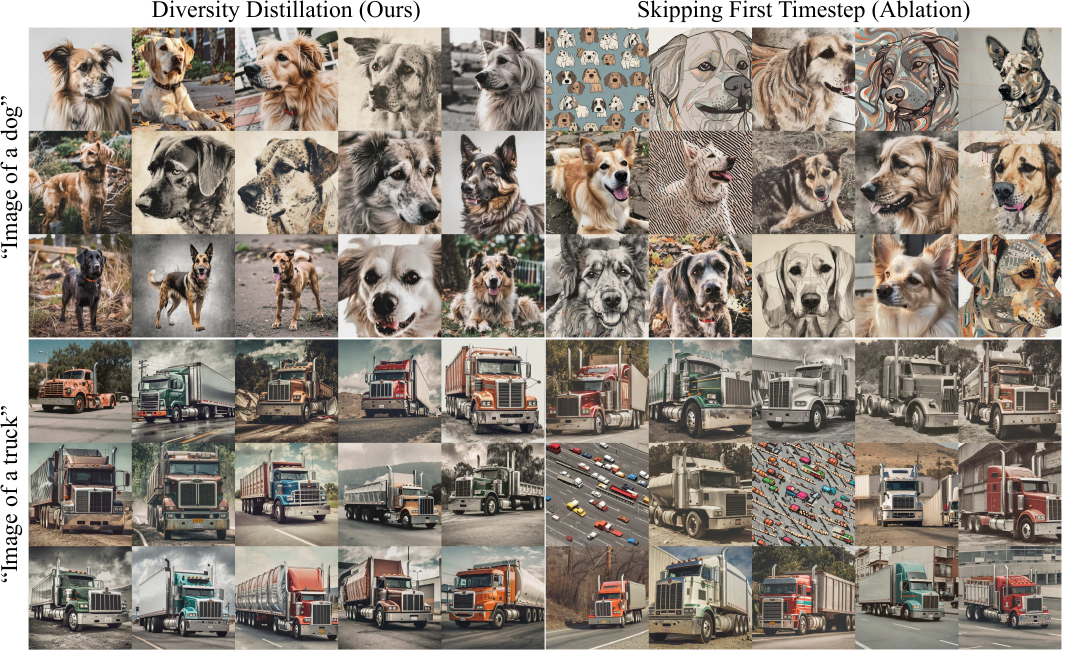}
\caption{Qualitative comparison between (left) our hybrid approach, (right) skip-first-step approach. The skip-first-step approach improves diversity over the standard distilled model but exhibits reduced quality compared to our hybrid method, particularly in fine details and coherence.}
\label{fig:skip_1}
\end{figure*}

In the main paper, we introduced a resource-efficient alternative to our hybrid approach: skipping the first timestep altogether in distilled model inference. We provide additional qualitative comparisons between this approach and our hybrid method in Figure~\ref{fig:skip_1}.

The skip-first-step approach provides a reasonable compromise when resource constraints prevent loading both models simultaneously. However, our quantitative analysis in the main paper and these qualitative examples demonstrate that the hybrid approach consistently achieves superior results in terms of both diversity and quality.

\section{Generalization Across Model Backbones}
\label{sec:other_models}
To assess the generality of our findings, we extend our analysis to different diffusion model architectures beyond SDXL. We evaluate two additional model pairs: PixArt-Alpha (base) with PixArt-Delta (distilled), and SD 2.1 (base) with SD-Turbo (distilled).

\begin{figure*}
\centering
\includegraphics[width=\textwidth]{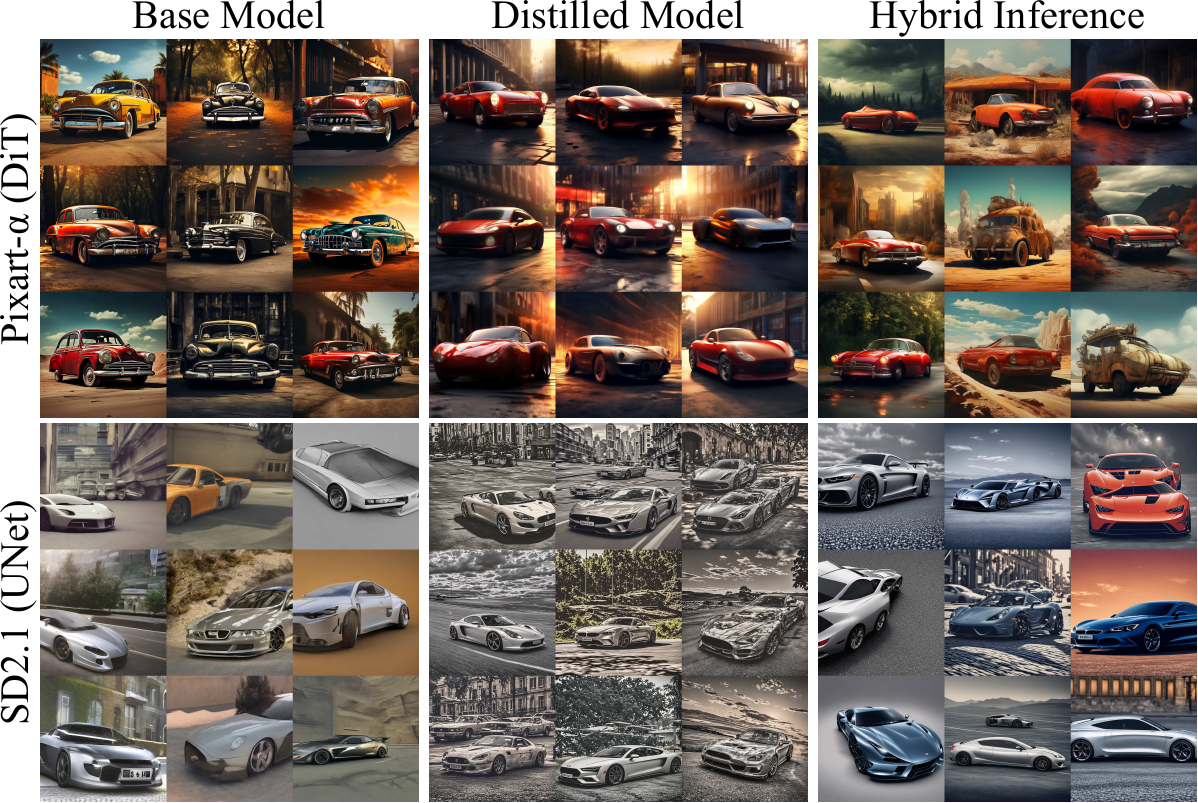}
\caption{\textbf{Generalization across model architectures.} Sample diversity comparison for the prompt "image of a car" across different diffusion model architectures. Top row: PixArt-$\alpha$ (DiT-based base model) shows diverse car types, colors, and contexts, while PixArt-$\delta$ (distilled) produces similar red sports cars with repetitive compositions. Bottom row: SD 2.1 (UNet-based base model) generates varied car styles and settings, while SD-Turbo (distilled) exhibits reduced diversity with similar silver/white cars in repetitive urban contexts. Our hybrid inference approach restores diversity in both architectures, demonstrating that the first-timestep bottleneck is architecture-agnostic.}
\label{fig:other_models}
\end{figure*}

Table~\ref{tab:generalization} shows that the diversity collapse phenomenon and the effectiveness of our solution generalize across different model architectures. PixArt-Delta exhibits similar sample diversity reduction compared to PixArt-Alpha, with our hybrid approach restoring diversity while maintaining efficiency. Similarly, SD-Turbo shows reduced sample diversity compared to SD 2.1, which our method successfully addresses. We show qualitative results in Figure~\ref{fig:other_models}

\begin{table}[h]
\centering
\resizebox{\linewidth}{!}{
\begin{tabular}{lcccc}
\toprule
Model & Architecture & Sample Diversity & Time (s) & Our Method \\
\midrule
PixArt-Alpha & DiT & 0.342 & 4.1 & - \\
PixArt-Delta & DiT & 0.198 & 0.9 & 0.339 \\
\midrule
SD 2.1 & UNet & 0.298 & 3.2 & - \\
SD-Turbo & UNet & 0.171 & 0.7 & 0.294 \\
\bottomrule
\end{tabular}
}
\caption{Sample diversity (DreamSim distance) across different model architectures for the prompt ``image of a car'' across 100 samples. Our hybrid approach consistently restores diversity regardless of the underlying architecture (DiT vs UNet) or distillation method. Please refer to Fig~\ref{fig:other_models} for qualitative samples}
\label{tab:generalization}
\end{table}

The consistent pattern across DiT-based (PixArt) and UNet-based (SD 2.1/Turbo) architectures demonstrates that the first-timestep diversity bottleneck is a fundamental characteristic of diffusion distillation, not specific to particular model designs.

\section{Causal Validation: Testing Later Timesteps}
\label{sec:causal}
To strengthen our causal claim that the first timestep is the critical bottleneck, we conduct a complementary experiment: replacing the \emph{final} timesteps of distilled models with base model steps while keeping the first timestep from the distilled model.

\begin{table}[h]
\centering
\resizebox{\linewidth}{!}{
\begin{tabular}{lcc}
\toprule
Method & Sample Diversity & Time (s) \\
\midrule
SDXL-Base & 0.357 & 9.22 \\
SDXL-DMD2 & 0.264 & 0.64 \\
\midrule
Replace First Timestep (Ours) & 0.350 & 0.64 \\
Replace Final Timesteps & 0.251 & 6.61 \\
\bottomrule
\end{tabular}
}
\caption{Causal validation experiment. Replacing final timesteps with base model steps provides minimal diversity improvement (DreamSim distance) compared to our first-timestep intervention, confirming that the first timestep is the critical bottleneck. The analysis is done for the prompt ``image of a car'' across 100 samples.}
\label{tab:causal_validation}
\end{table}

\begin{figure*}
\centering
\includegraphics[width=\textwidth]{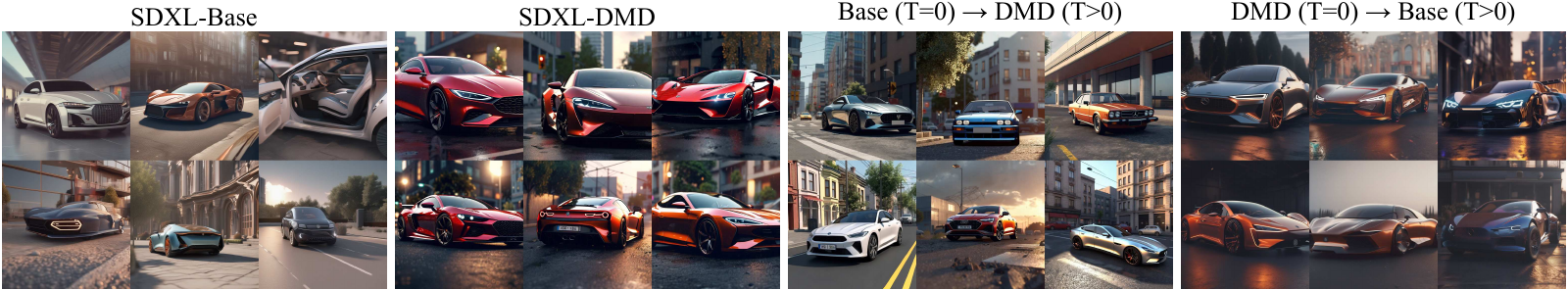}
\caption{\textbf{Causal validation of first-timestep importance.} Visual comparison for the prompt "image of a car" showing: (left) SDXL-Base with diverse car types, colors, and contexts; (middle-left) SDXL-DMD with reduced diversity showing similar red sports cars; (middle-right) our hybrid approach using base model for first timestep (T=0) then DMD for remaining steps, successfully restoring diversity; (right) control experiment using DMD for first timestep (T=0) then base model for remaining steps, showing minimal diversity improvement. This demonstrates that the first timestep, not later steps, controls sample diversity.}
\label{fig:causal}
\end{figure*}

Table~\ref{tab:causal_validation} shows that replacing the final timesteps does not yield diversity improvements compared to our first-timestep intervention. This result provides strong causal evidence that the diversity bottleneck is concentrated at the beginning of the generation process, not distributed throughout the timesteps. The minimal improvement from modifying later steps confirms our $\hat{\mathbf{x}}_{0}$ analysis: once the structural decisions are made in the first timestep, later steps primarily refine details rather than introduce fundamental variations. We show qualitative examples in Figure~\ref{fig:causal}

\section{Extended \texorpdfstring{$\hat{\mathbf{x}}_{0}$}{x̂₀} visualization Analysis}
\label{sec:more_x0}
\begin{figure*}
\centering
\includegraphics[width=\textwidth]{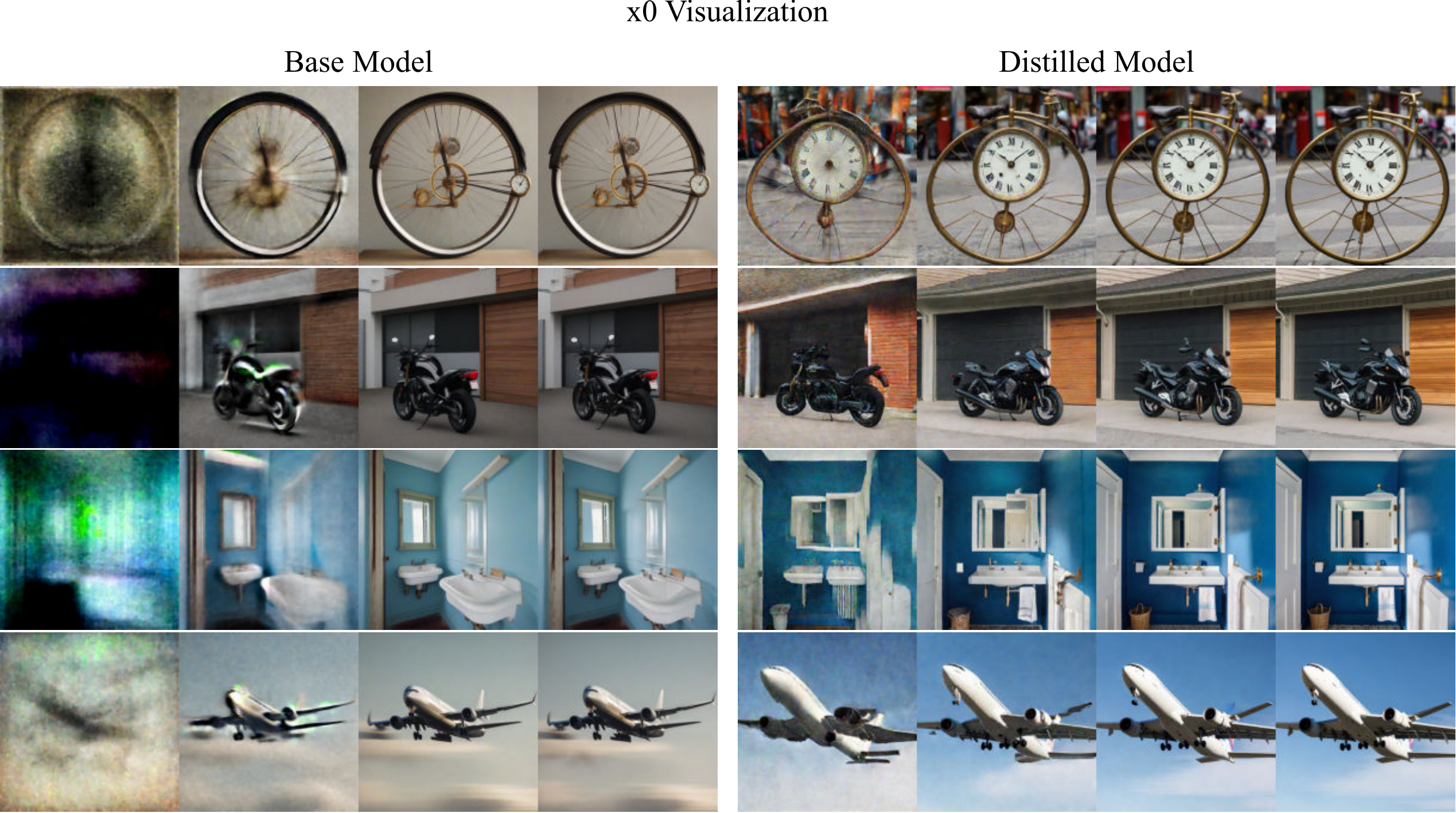}
\caption{Extended $\hat{\mathbf{x}}_{0}$ visualization comparison between SDXL-Base and SDXL-DMD for the prompt. The visualization reveals that DMD commits to final structural composition within the first timestep, while Base gradually develops structure across multiple steps. This pattern is consistent across different content types and prompts.}
\label{fig:dt_1}
\end{figure*}

\begin{figure*}
\centering
\includegraphics[width=\textwidth]{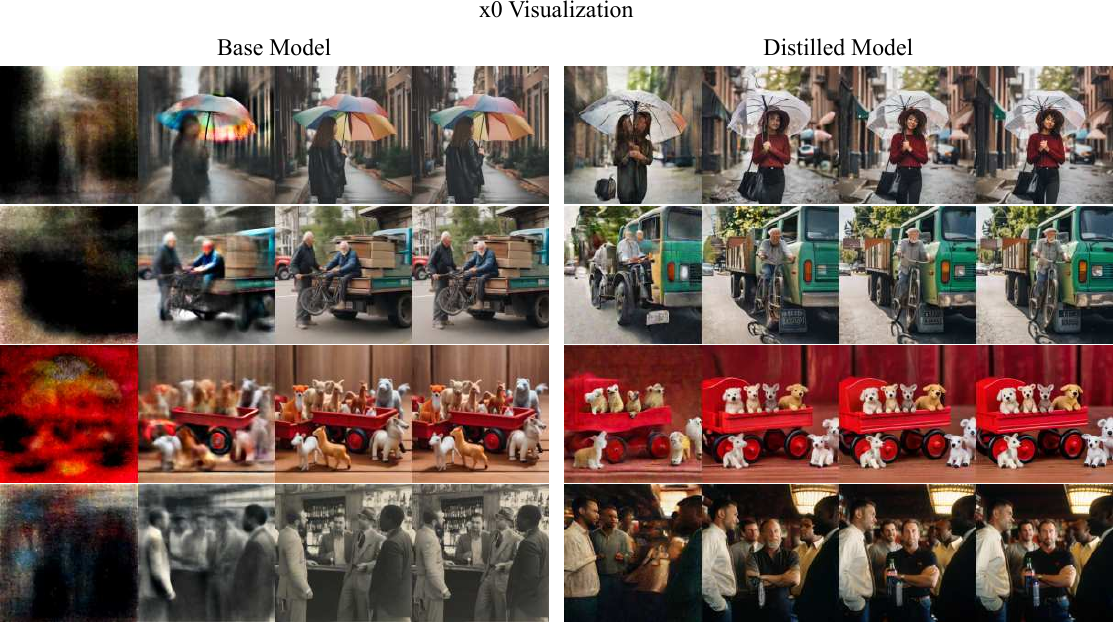}
\caption{Extended $\hat{\mathbf{x}}_{0}$ visualization comparison between SDXL-Base and SDXL-DMD for the prompt. The visualization reveals that DMD commits to final structural composition within the first timestep, while Base gradually develops structure across multiple steps. This pattern is consistent across different content types and prompts}
\label{fig:dt_2}
\end{figure*}

The main paper introduced $\hat{\mathbf{x}}_{0}$ visualization technique for analyzing how diffusion models develop structural information during the denoising process. We present additional visualizations in Figures~\ref{fig:dt_1},~\ref{fig:dt_2} that further illuminate the differences between base and distilled models.
 
These visualizations reinforce our key finding: distilled models compress the diversity-generating behavior distributed across early timesteps in base models into a single initial step, explaining the observed mode collapse. This insight directly informed our hybrid inference approach, which strategically leverages the diversity-generating capabilities of base models in critical early steps.

\section{Mode Collapse and Diversity}
\label{sec:more_examples}
\begin{figure*}
\centering
\includegraphics[width=\textwidth]{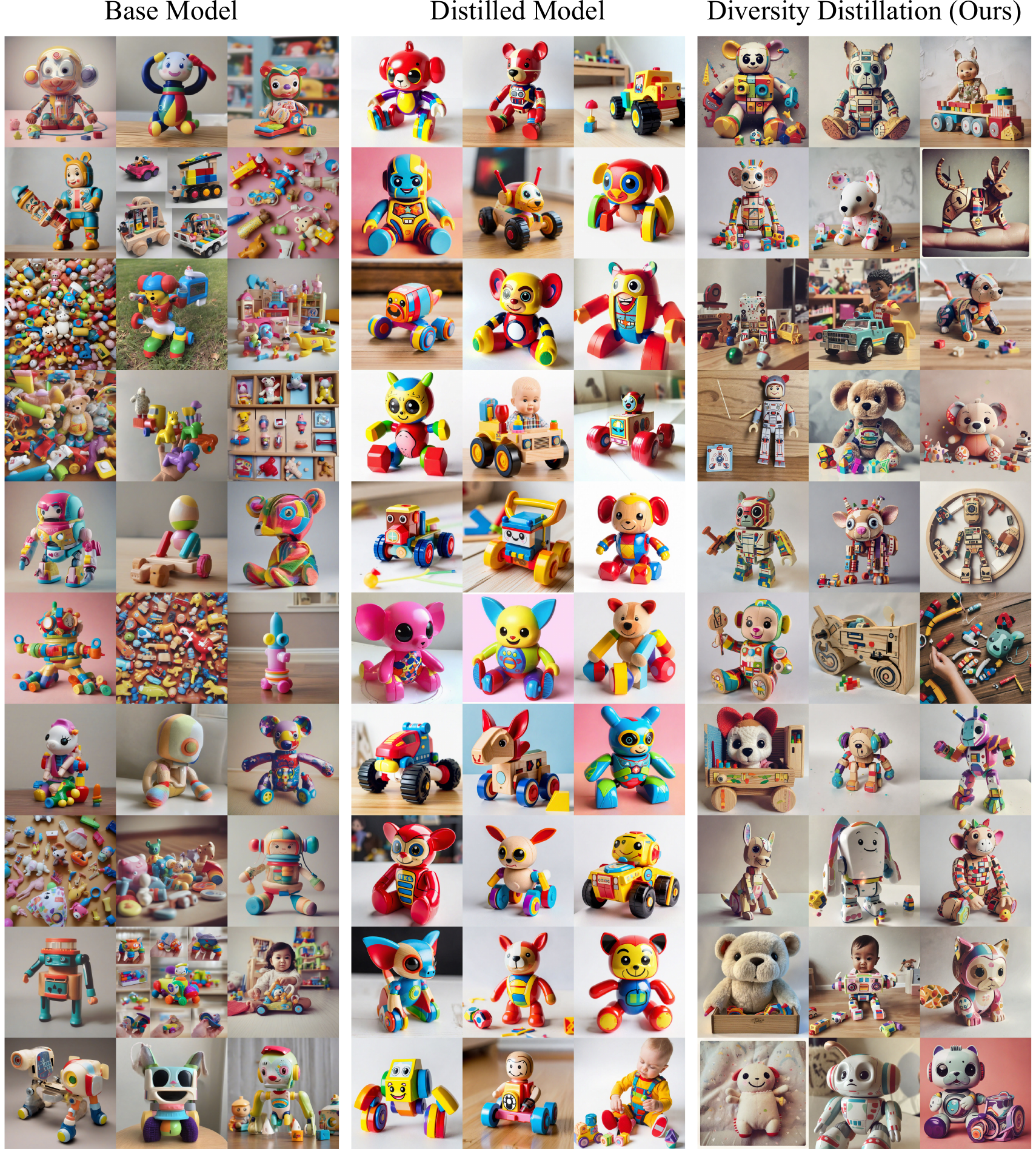}
\caption{Comparison of generation diversity across different models for the prompt "image of a toy." Each image shows different seeds for the same model. Note the structural similarity in distilled model outputs compared to the greater variation in base model and our hybrid approach.}
\label{fig:diversity_1}
\end{figure*}

\begin{figure*}
\centering
\includegraphics[width=\textwidth]{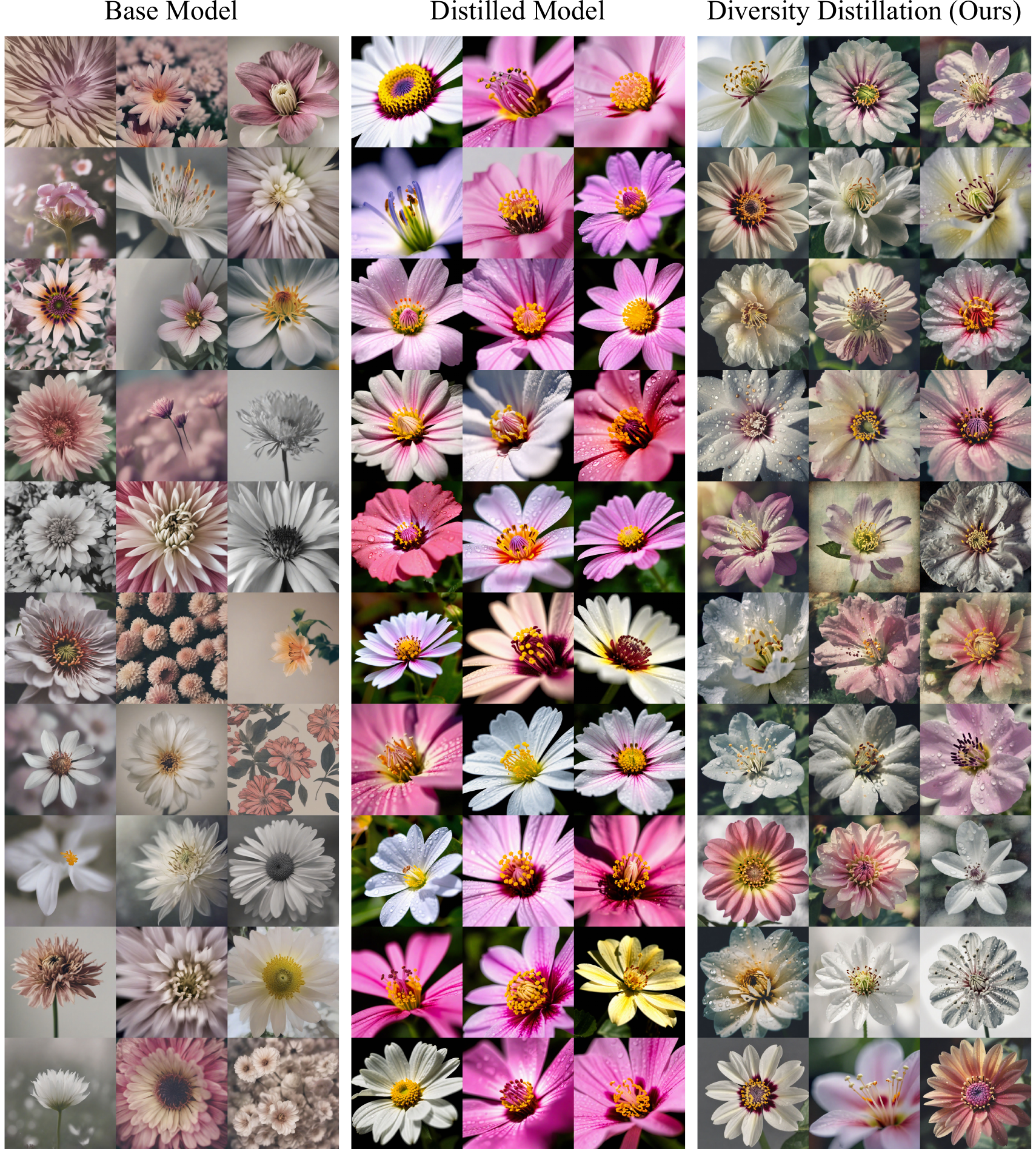}
\caption{Comparison of generation diversity for "image of a flower" Distilled models (middle column) produce structurally similar outputs across different seeds, while our approach (right column) restores diversity comparable to the base model (left column) while maintaining the speed advantage of distilled models.}
\label{fig:diversity_2}
\end{figure*}

\begin{figure*}
\centering
\includegraphics[width=\textwidth]{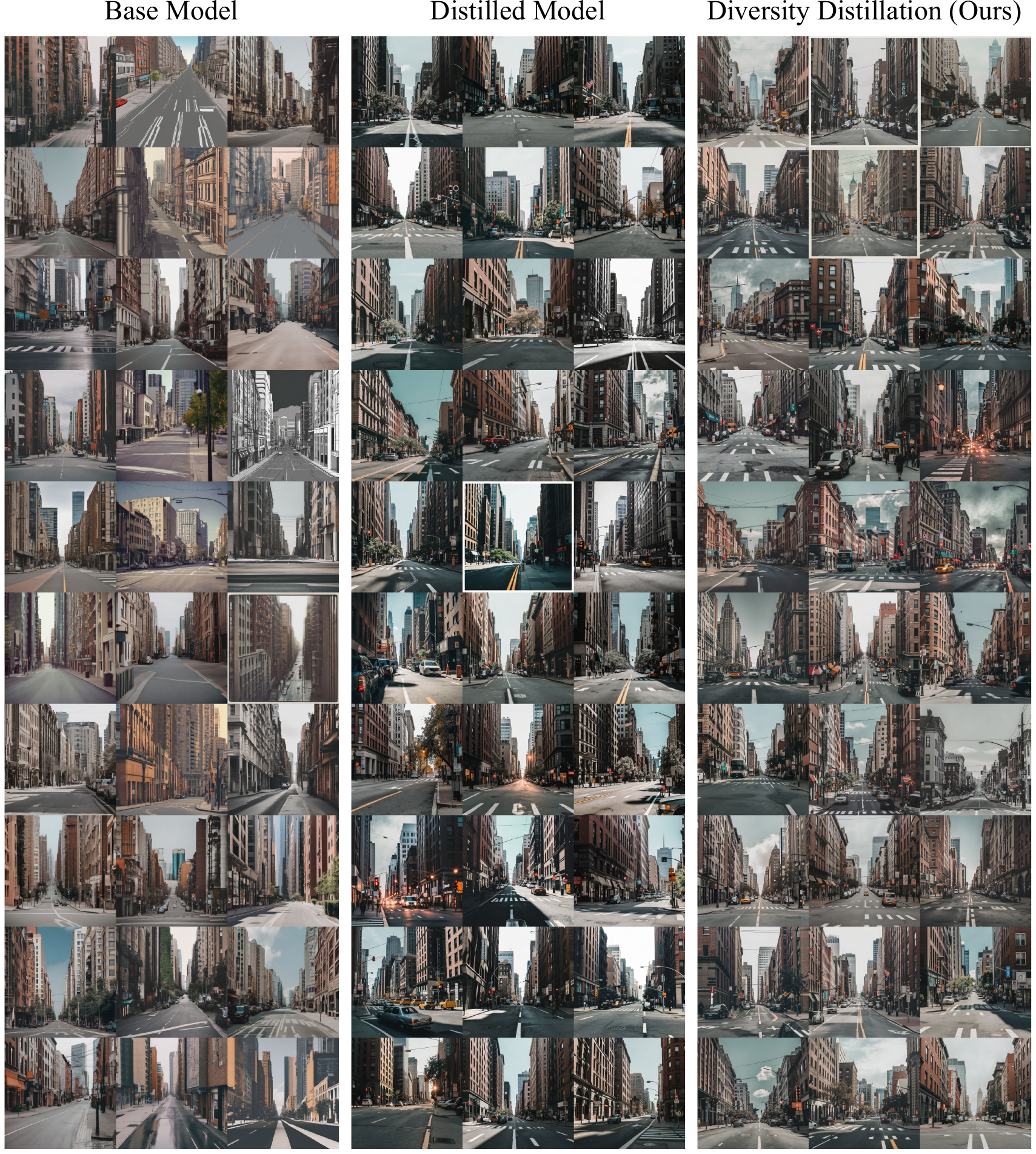}
\caption{Additional diversity comparison for "city street"Distilled models (middle column) produce structurally similar outputs across different seeds, while our approach (right column) restores diversity comparable to the base model (left column) while maintaining the speed advantage of distilled models.}
\label{fig:diversity_3}
\end{figure*}

\begin{figure*}
\centering
\includegraphics[width=\textwidth]{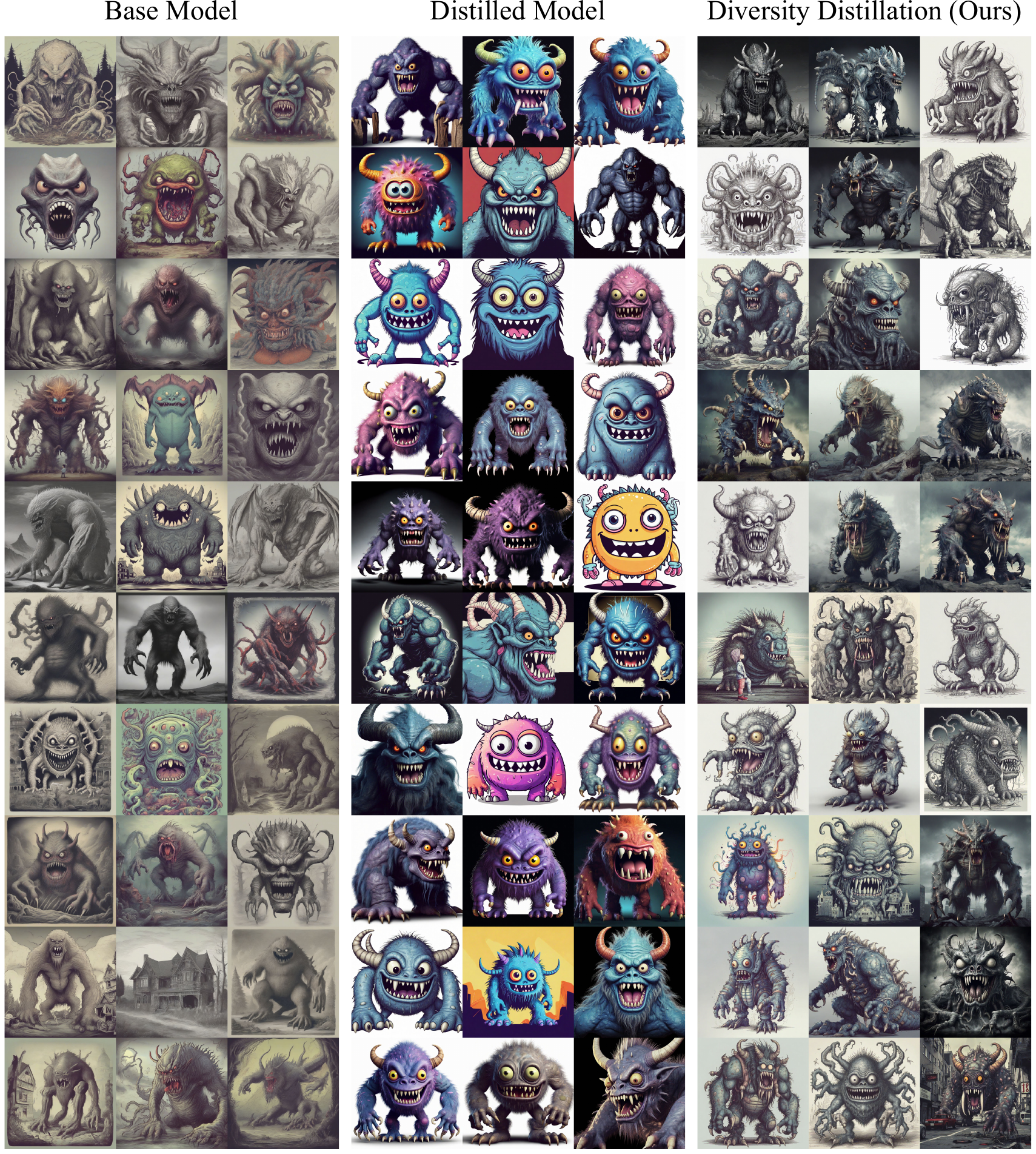}
\caption{Diversity comparison for abstract prompt: "picture of a monster" Distilled models (middle column) produce structurally similar outputs across different seeds, while our approach (right column) restores diversity comparable to the base model (left column) while maintaining the speed advantage of distilled models.}
\label{fig:diversity_4}
\end{figure*}

The main paper introduced our finding that distilled diffusion models suffer from reduced sample diversity (mode collapse) compared to their base counterparts. We provide additional qualitative examples in Figure~\ref{fig:diversity_1}-\ref{fig:diversity_4} that visually demonstrate this phenomenon across various prompts and model variants.

These examples highlight the significant diversity loss in distilled models. While the distilled models produce high-quality images, they often converge to similar structural compositions regardless of random seed initialization. Our diversity distillation approach effectively addresses this limitation, restoring the variety of outputs comparable to the base model while maintaining computational efficiency.

\end{document}